\documentclass{PoS}

\newcommand{\Tr}{\mbox{Tr}}

\title{The equation of state with nonzero chemical potential for 2+1 flavors}

\ShortTitle{The equation of state with nonzero chemical potential for 2+1 flavors}

\author{C.~Bernard\\
        Physics Department, Washington University, St. Louis, MO 63130, USA}
\author{T.~Burch\\
        Institut f\"ur Theoretische Physik, Universit\"at Regensburg, D-93040 Regensburg, Germany}
\author{C.~DeTar and L.~Levkova\\
        Physics Department, University of Utah, Salt Lake City, UT 84112, USA}
\author{Steven~Gottlieb\\
        Physics Department, Indiana University, Bloomington, IN 47405, USA}
\author{\speaker{U.~M.~Heller}\\
        American Physical Society, One Research Road, Box 9000, Ridge, NY 11961-9000, USA\\
        E-mail: \email{heller@aps.org}}
\author{J.~E.~Hetrick\\
        Physics Department, University of the Pacific, Stockton, CA 95211, USA}
\author{R.~Sugar\\
        Physics Department, University of California, Santa Barbara, CA 93106, USA}
\author{D.~Toussaint\\
        Physics Department, University of Arizona, Tucson, AZ 85721, USA}

\abstract{
We present results for the QCD equation of state with nonzero
chemical potential using the Taylor expansion method with terms
up to sixth order in the expansion. Our calculations are
performed on asqtad 2+1 quark flavor lattices at $N_t=4$.
}

\FullConference{The XXV International Symposium on Lattice Field Theory\\
		 July 30-4 August 2007\\
		 Regensburg, Germany}

\begin{document}

\section{Introduction}

The cost of the computation of the equation of state (EoS) increases
very quickly with increasing temporal extent $N_t$ and aspect ratio
$N_s/N_t$ fixed. Since the lattice spacing for small $N_t$
-- $a=1/(T N_t)$ -- can be quite large, especially at low temperatures,
it is very important to use improved actions with small
discretization errors. The MILC collaboration has a longstanding program
of full $2+1$ flavor simulations using the asqtad quark action \cite{asq}
combined with a one-loop Symanzik improved gauge action \cite{sym}.
This includes a computation of the EoS at zero baryon chemical potential
and hence vanishing baryon density \cite{MILC_EoS}. Here we will present
the results of an extension of this computation to better approximate
conditions in heavy ion collision experiments, namely the inclusion of
a (small) chemical potential.

To avoid the notorious sign problem -- the fermion determinant becomes
complex with a nonzero chemical potential making straight forward
Monte Carlo simulations impossible -- we use the Taylor
expansion method \cite{taylor} which requires only simulations at
zero chemical potential and only on the finite temperature ensembles.

\section{Technicalities}

In the Taylor expansion method \cite{taylor}
one expands, for example, the pressure as
\begin{equation}
\frac{p}{T^4}=\frac{\ln Z} {VT^3}=
\sum_{n,m=0}^\infty c_{nm}(T) \left(\frac{\bar{\mu}_l}{T}\right)^n
\left(\frac{\bar{\mu}_h}{T}\right)^m.\label{eq:pmu}
\end{equation}
Here $Z$ is the partition function, and $\bar{\mu}_{l,h}$ are the
chemical potentials for the light and heavy quarks, respectively.
The expansion coefficients are evaluated at zero chemical potential
$\mu_{l,h}=0$.
Due to CP symmetry the terms in the series with $n+m$ odd vanish.
The nonzero coefficients are
\begin{equation}
c_{nm} (T)=
\frac{1}{n!}\frac{1}{m!}\frac{N_t^{3}}{ N_s^3}\frac{\partial^{n+m}\ln Z}
{\partial(\mu_l N_t)^n\partial(\mu_h N_t)^m}\biggr\vert_{\mu_{l,h}=0} \quad,
\label{eq:cn}
\end{equation}
where now the $\mu_{l,h}$ are the chemical potentials in lattice units. 
Knowledge of the $c_{nm}(T)$ also allows for the computation of quark
number densities and susceptibilities from eq.~(\ref{eq:pmu}), since
\begin{equation}
\frac{n_j}{T^3} = \frac{\partial}{\partial \bar{\mu}_j/T}
 \left(\frac{\ln Z}{T^3V}\right) , \qquad\qquad {\rm for~}j=l,h
\end{equation}
and
\begin{equation}
\frac{\chi_{ij}}{T^2} = \frac{\partial}{\partial \bar{\mu}_i/T}
 \left(\frac{n_j}{T^3}\right) , \qquad\qquad {\rm for~}i,j=l,h .
\end{equation}

Similarly, the interaction measure is expanded as
\begin{equation}
\frac{I}{T^4}=-\frac{N_t^3}{N_s^3} \frac{d\ln Z}{d\ln a}=
 \sum_{n,m=0}^\infty b_{nm}(T)
 \left(\frac{\bar{\mu}_l}{T}\right)^n \left(\frac{\bar{\mu}_h}{T}\right)^m,
\end{equation}
where again only terms with $n+m$ even are nonzero and
\begin{equation}
b_{nm}(T) = \left.-\frac{1}{n!}\frac{1}{m!}\frac{N_t^3}{N_s^3} 
 \frac{\partial^{n+m}}{\partial(\mu_l N_t)^n\partial(\mu_h N_t)^m}
 \right|_{\mu_{l,h}=0}\left(\frac{d\ln Z}{d\ln a} \right).
\end{equation}
For explicit expressions of the Taylor expansion coefficients $c_{nm}(T)$
and $b_{nm}(T)$ we refer the reader to \cite{nonzero_mu}.

To determine the Taylor expansion coefficients $c_{nm}(T)$ and $b_{nm}(T)$
in numerical simulations, we need to calculate traces of derivatives of
the asqtad fermion matrix such as
\begin{equation}
\frac{\partial^{n} \ln \det M_{l,h}}{\partial \mu_{l,h}^n}
 = \frac{\partial^{n} \,\Tr \ln M_{l,h}}{\partial \mu_{l,h}^n}, \hspace{1cm}
\frac{\partial^{n} \,\Tr M_{l,h}^{-1}}{\partial \mu_{l,h}^n}, \hspace{1cm}
\frac{\partial^{n} \,\Tr (M_{l,h}^{-1}\frac{dM_{l,h}}{du_0})}{\partial \mu_{l,h}^n},
\end{equation}
and products of such traces. The number of such terms increases fast
with increasing order $n+m$. Up to sixth order 40 such terms need to be
computed.
These traces are estimated on the ensembles of lattices along a
trajectory of constant physics using 200 random sources in the region 
of the phase transition/crossover and 100 sources outside that region.
With these numbers of random sources the noise in the Taylor expansion
coefficients is dominated by configuration-to-configuration fluctuations.
Increasing the number of random sources would thus not decrease our
statistical errors substantially.

\section{First numerical results}

\begin{figure}[h]
\begin{center} 
\includegraphics[width=12cm]{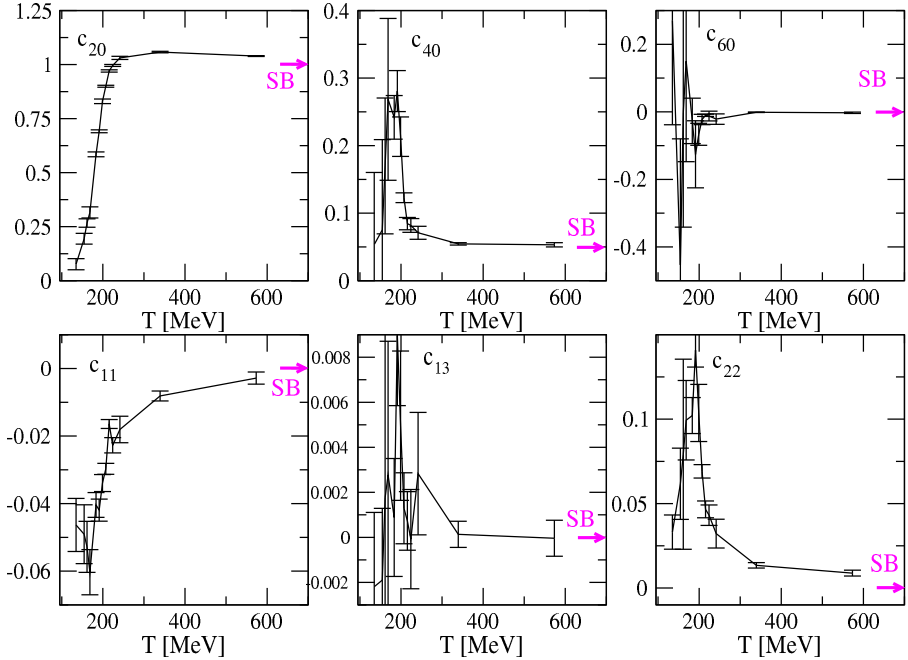}
\end{center}\vspace{-0.3cm} 
\caption{Some nonvanishing coefficients in the Taylor expansion of the
pressure as function of the temperature along the trajectory of
constant physics.}
\label{fig:Cnm}
\end{figure}

We have computed the Taylor expansion coefficients up to sixth order
on the lattice ensembles along one trajectory of constant physics where
we had computed the EoS at vanishing chemical potential before
\cite{MILC_EoS}. The trajectory is given by the choices $N_t=4$ and
$m_{ud} \approx 0.1 m_s$, where $m_s$ is tuned to the physical strange
quark mass to within about 20\%.

\begin{figure}[h]
\begin{center} 
\includegraphics[width=12cm]{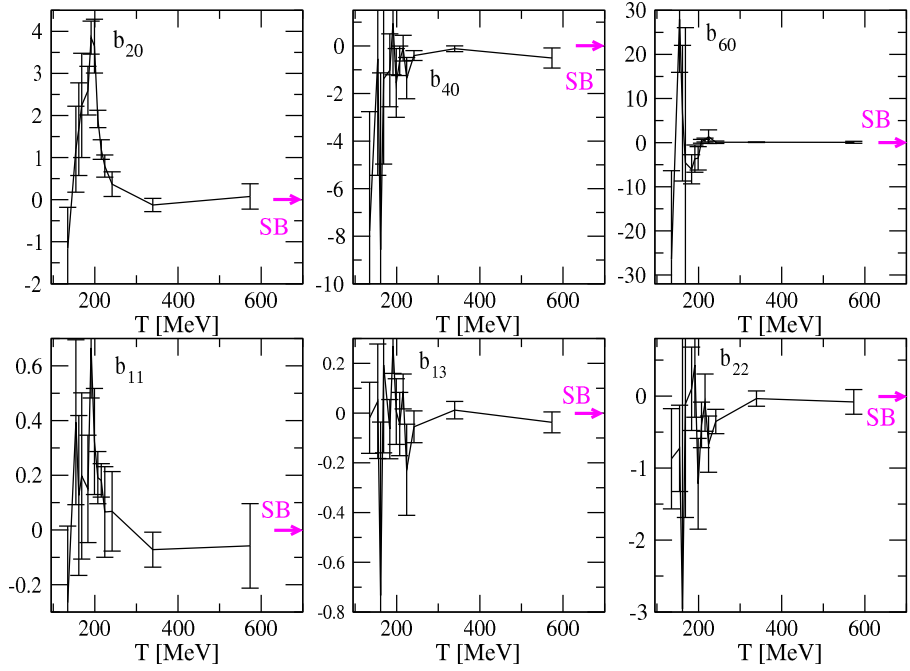}
\end{center}\vspace{-0.3cm} 
\caption{Some nonvanishing Taylor coefficients in the expansion of the
interaction measure.}
\label{fig:Bnm}
\end{figure}

\begin{figure}[h]
\vspace{0.3cm}
\begin{minipage}{17pc}
\includegraphics[width=16pc]{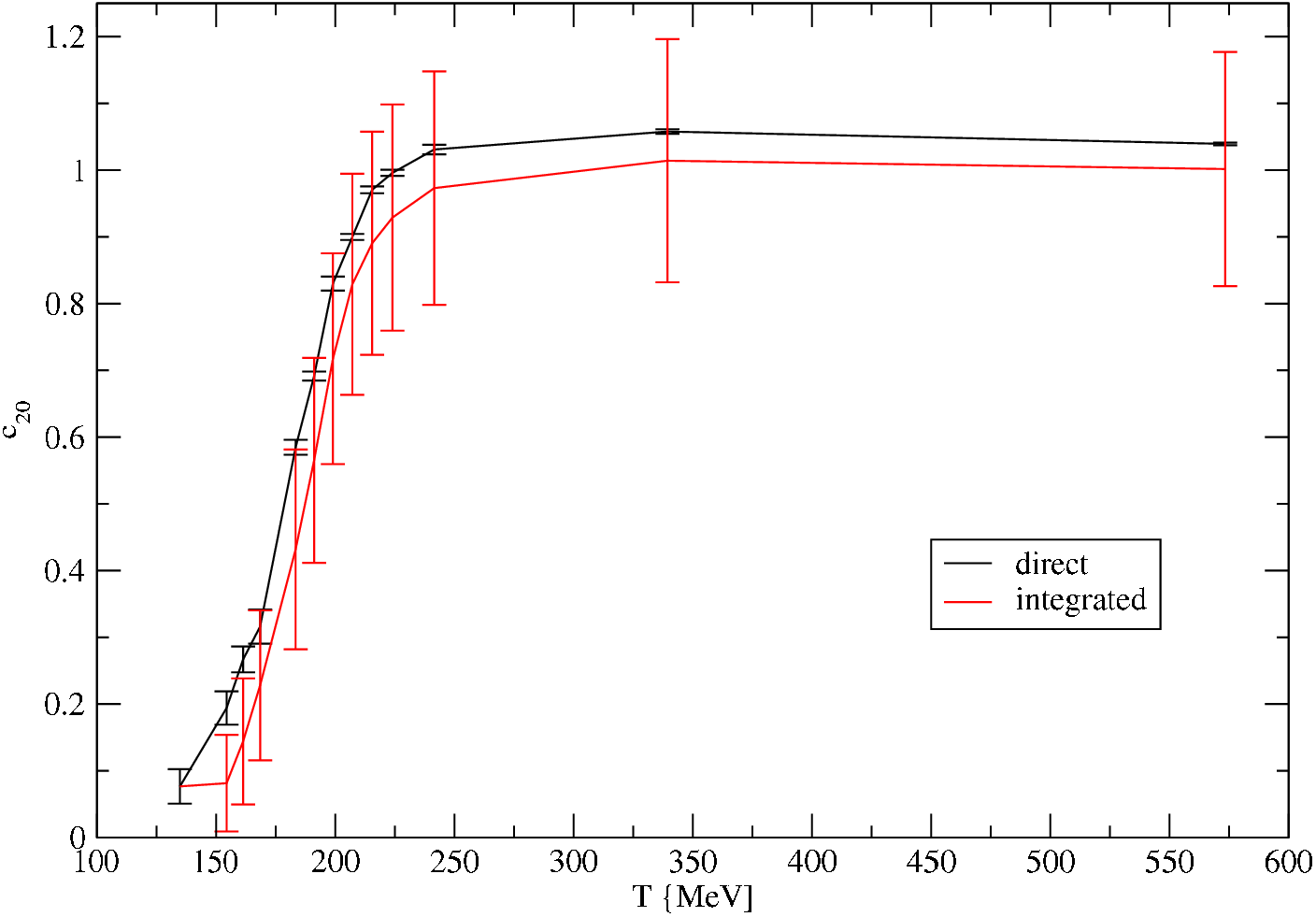}
\label{fig:C20comp}
\caption{Comparing the two methods to compute $c_{20}(T)$, directly or by
integrating $b_{20}(T)$.}
\end{minipage}\hspace{2pc}%
\begin{minipage}{17pc}
\includegraphics[width=16pc]{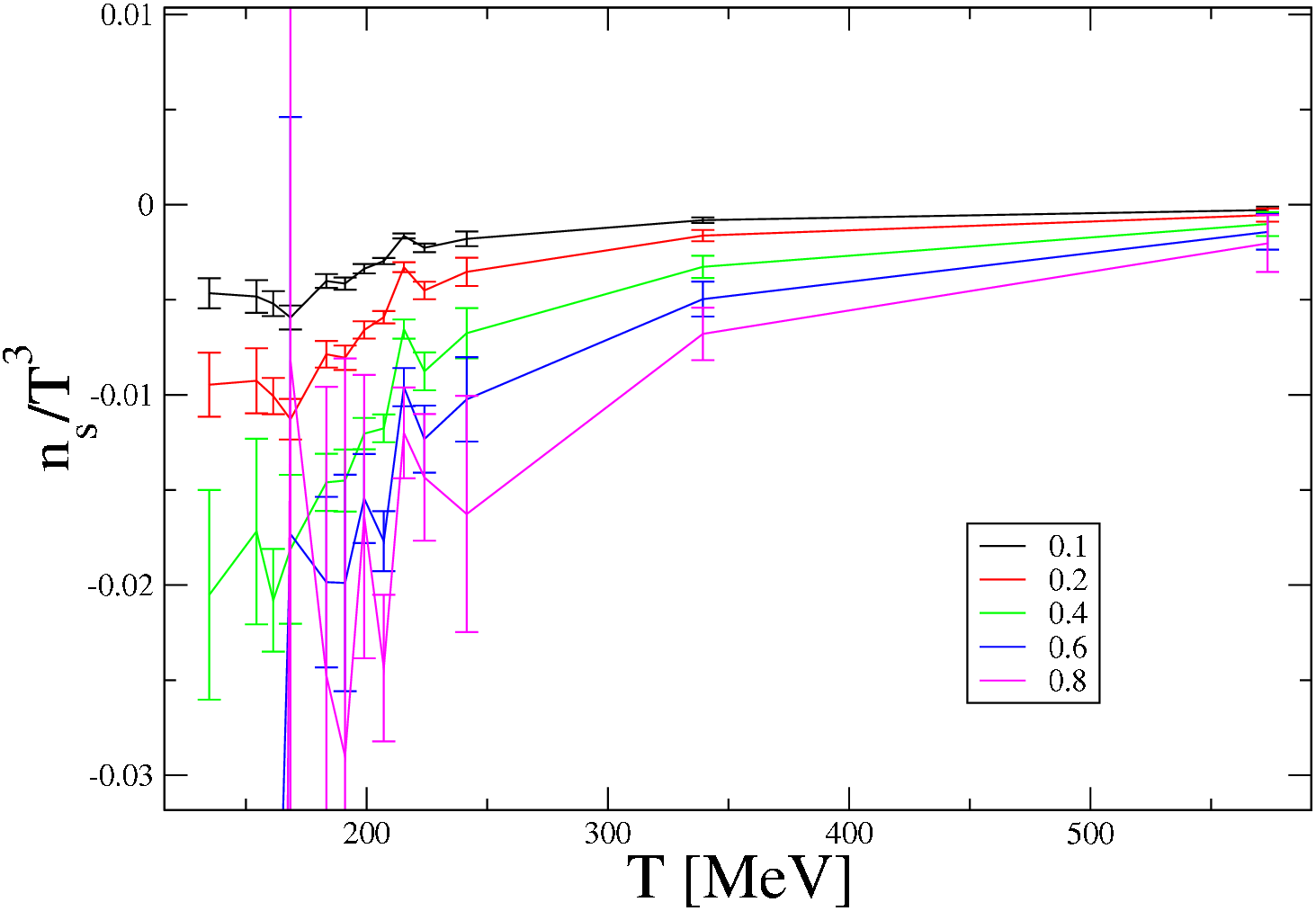}
\caption{The induced strange quark number density $n_s/T^3$ when turning on
only $\bar{\mu}_l / T$ and keeping $\bar{\mu}_h / T = 0$.}
\label{fig:Ns_untunes}
\end{minipage}\hspace{2pc}%
\end{figure}

Figure~\ref{fig:Cnm} shows our result for some of the coefficients
for the contribution to the pressure difference
$\Delta p = p(\mu_{l,h}\neq0) - p(\mu_{l,h}=0)$, see eq.~(\ref{eq:pmu}).
Note that the coefficients quickly reach the continuum Stefan-Boltzmann
limit above $T_c$. Also note that the mixed coefficients with both
$n,m \ne 0$ are quite small.
Some coefficients contributing to the difference for the interaction
measure $\Delta I = I(\mu_{l,h}\neq0) - I(\mu_{l,h}=0)$ are show in
figure~\ref{fig:Bnm}.
Since the pressure can be obtained by integration of the interaction
measure along the trajectory of constant physics, $a^4 p = \int_{TCP}
a^{\prime 4} I(a^\prime) d \ln a^\prime$, the coefficients $c_{nm}(T)$
can be obtained by integrating $b_{nm}(T)$, giving a consistency check.
This is illustrated in figure~\ref{fig:C20comp}.

\begin{figure}[h]
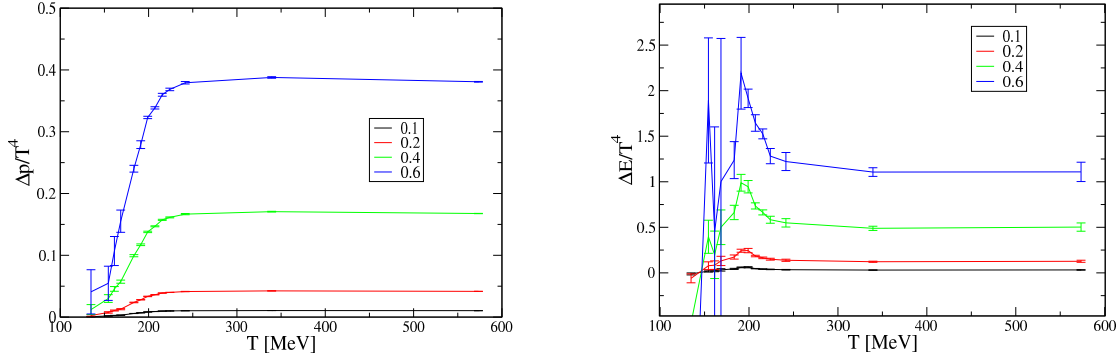

\begin{minipage}{17pc}
\includegraphics[width=16pc]{dPO6.eps}
\end{minipage}\hspace{2pc}%
\begin{minipage}{17pc}
\includegraphics[width=16pc]{dEO6.eps}
\end{minipage}\hspace{2pc}%
\caption{Pressure, $\Delta p$, and energy density, $\Delta \varepsilon$,
contribution for the $m_{ud}\approx 0.1 m_s$, $N_t=4$ trajectory
to $O(\mu^6)$.}
\label{fig:EOSmu}
\end{figure}

Turning on $\bar{\mu}_l / T$ induces a small negative strange quark number
density $n_s/T^3$ because some of the $c_{n1}(T)$ are nonvanishing, as
shown in figure~\ref{fig:Ns_untunes}. To keep $n_s =0$, as is the case
in heavy ion experiments, we need to compensate by a small tuned
$\bar{\mu}_h / T$. We show the pressure and energy density contribution
due to the nonzero chemical potential, with tuned $n_s = 0$, in
Figure~\ref{fig:EOSmu}.

\section{The isentropic equation of state}

\begin{figure}[h]
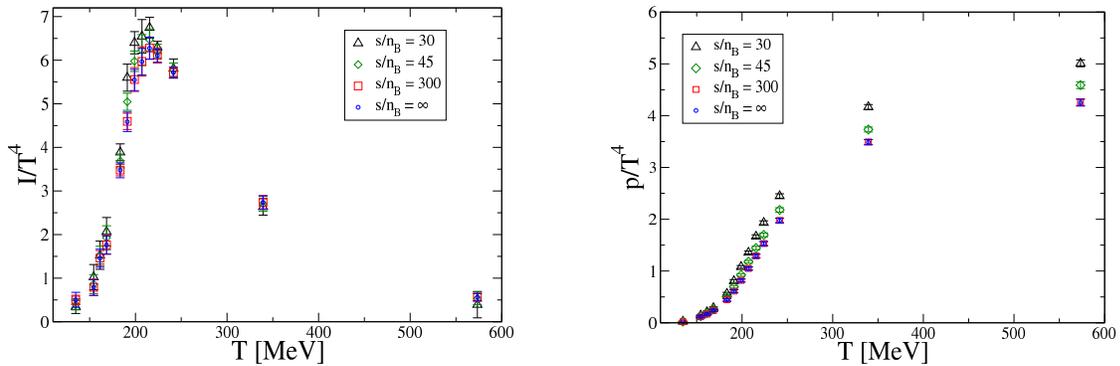

\begin{minipage}{17pc}
\includegraphics[width=16pc]{IO6Isentr.eps}
\end{minipage}\hspace{2pc}%
\begin{minipage}{17pc}
\includegraphics[width=16pc]{PO6Isentr.eps}
\end{minipage}\hspace{2pc}%
\caption{Isentropic versions of the interaction measure (left) and
pressure (right).}
\label{fig:Isentr_1}
\end{figure}

\begin{figure}[h]
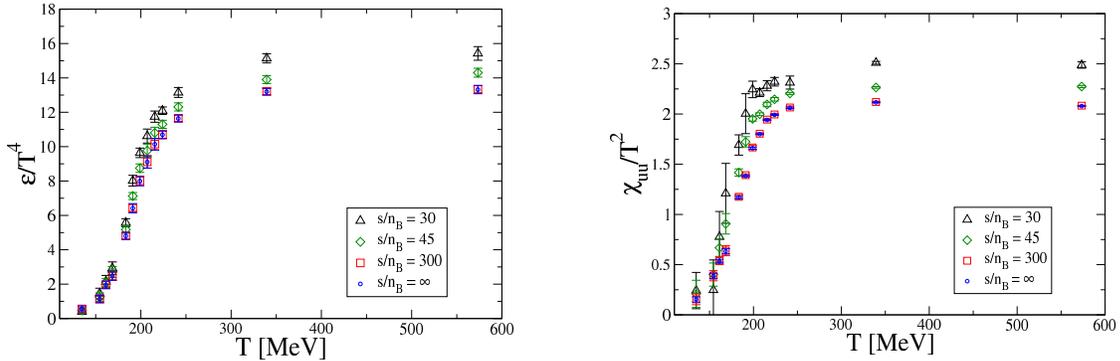

\begin{minipage}{17pc}
\includegraphics[width=16pc]{EO6Isentr.eps}
\end{minipage}\hspace{2pc}%
\begin{minipage}{17pc}
\includegraphics[width=16pc]{Chi_uuO6Isentr.eps}
\end{minipage}\hspace{2pc}%
\caption{The isentropic energy density (left) and light-light quark number
susceptibility (right).}
\label{fig:Isentr_2}
\end{figure}

Heavy ion collision experiments produce matter that, after thermalization,
is expected to expand, at fixed baryon number, without further entropy
generation, {\it i.e.} isentropically -- with constant $s/n_B$.
At the AGS, SPS and RHIC $s/n_B$ is approximately 30, 45 and 300
\cite{Ejiri:2005uv}, respectively. To account for this situation,
we numerically determine the $\mu_l$ and $\mu_h$, as function of $T$,
by solving
\begin{equation}
\frac{s}{n_B}(\mu_l, \mu_h) = C , \qquad
\frac{n_s}{T^3} (\mu_l, \mu_h) = 0,
\end{equation}
for $C=30$, 40, and 300, within our statistical errors. With the
determined $\mu_l$ and $\mu_h$ we can then compute the isentropic
equation of state, shown in figures~\ref{fig:Isentr_1} and
\ref{fig:Isentr_2} (left). For comparison, we also include the results
with $\mu_l=\mu_h=0$, {\it i.e.} $s/n_B=\infty$.

In figure~\ref{fig:Isentr_2} (right) we show, as a further example,
the isentropic light-light quark number susceptibility, $\chi_{uu}$.
We note that $\chi_{uu}$ does not develop a peak structure on any of the
isentropic trajectories, as would be expected near a phase transition
point. Therefore all the mentioned heavy ion experiments
take place far away from a possible critical (end) point in the $\mu$ -- $T$
plane of the phase diagram.

\section{Conclusions}

We have extended the computation of the QCD equation of state for
$2+1$ flavors along a trajectory of constant physics with $m_{ud}/m_s
\approx 0.1$ on lattice ensembles with $N_t=4$ to small nonzero
chemical potential with the Taylor expansion
method up to sixth order in the chemical potential. We tuned the
strange quark chemical potential to keep the strange quark density
vanishing at different values of the light quark chemical potential.

We have also determined the isentropic EoS and quark number susceptibilities
for values of the ratio $s/n_B$ relevant for heavy ion collision
experiments. We found no signs of a possible phase transition along
any of the considered isentropic trajectories. Qualitatively our results
are in agreement with previous two-flavor studies.

\begin{center} 
{\bf Acknowledgments}\\
\end{center}
This work was supported by the US DOE and NSF.
Computations were performed at CHPC (Utah), FNAL, FSU, IU, NCSA and UCSB.

\end{document}